\documentclass[11pt,twocolumn,tightenlines,superscriptaddress]{revtex4}
\usepackage[latin9]{inputenc}
\usepackage{bm}
\usepackage{amsmath}
\usepackage{amssymb}
\usepackage{graphicx}
\usepackage{esint}
\usepackage{comment}
\usepackage{natbib}

\makeatletter
\@ifundefined{textcolor}{}
{%
 \definecolor{BLACK}{gray}{0}
 \definecolor{WHITE}{gray}{1}
 \definecolor{RED}{rgb}{1,0,0}
 \definecolor{GREEN}{rgb}{0,1,0}
 \definecolor{BLUE}{rgb}{0,0,1}
 \definecolor{CYAN}{cmyk}{1,0,0,0}
 \definecolor{MAGENTA}{cmyk}{0,1,0,0}
 \definecolor{YELLOW}{cmyk}{0,0,1,0}
}

\newcommand{\ket}[1]{\left| {#1} \right\rangle}
\newcommand{\bra}[1]{\left\langle {#1}\right|}
\newcommand{\braket}[1]{\langle {#1} \rangle}
\newcommand{\abs}[1]{\left|| \,{#1} \right|\!|}
\newcommand{\ve}[1]{\mathbf{#1}}
\newcommand{\com}[2]{[ {#1}\, ,{#2} ]}
\newcommand{\acom}[2]{\left\{ {#1}\, ,{#2} \right\}}

\newcommand{\Tr}{\text{Tr}}

\newcommand{\s}{\text{\tiny{$\mathcal{S}$}}}

\newcommand{\I}{\text{\tiny{I}}}
\newcommand{\R}{\text{\tiny{R}}}

\newcommand{\st}{h} 
\newcommand{\av}[1]{\mathbb{E}[ #1 ]}

\usepackage{color}\usepackage{amsthm}\usepackage{bbm}
\usepackage{epsfig}\usepackage{lscape}\usepackage{float}\usepackage{subfigure}\usepackage{dcolumn}\usepackage{bbm}\usepackage{color}\usepackage{epstopdf}\usepackage{amscd}\usepackage{amsfonts}\usepackage{mathrsfs}\usepackage{cases}

\makeatother

\begin{document}

\title{Gravity induced wave function collapse}

\author{G. Gasbarri}

\affiliation{Abdus Salam ICTP, Strada Costiera 11, I-34151 Trieste, Italy} 

\affiliation{Department of Physics, University of Trieste, Strada Costiera 11,
34151 Trieste, Italy}

\affiliation{Istituto Nazionale di Fisica Nucleare, Trieste Section, Via Valerio
2, 34127 Trieste, Italy}

\author{M. Toro\v{s}}

\affiliation{Department of Physics, University of Trieste, Strada Costiera 11,
34151 Trieste, Italy}

\affiliation{Istituto Nazionale di Fisica Nucleare, Trieste Section, Via Valerio
2, 34127 Trieste, Italy}

\author{S. Donadi}

\affiliation{Department of Physics, University of Trieste, Strada Costiera 11,
34151 Trieste, Italy}

\affiliation{Istituto Nazionale di Fisica Nucleare, Trieste Section, Via Valerio
2, 34127 Trieste, Italy}

\author{A. Bassi}

\affiliation{Department of Physics, University of Trieste, Strada Costiera 11,
34151 Trieste, Italy}

\affiliation{Istituto Nazionale di Fisica Nucleare, Trieste Section, Via Valerio
2, 34127 Trieste, Italy}
\begin{abstract}
Starting from an idea of S.L. Adler~\cite{Adler2015}, we develop a novel model of gravity-induced spontaneous wave-function collapse. The collapse is driven by complex stochastic fluctuations of the spacetime metric. After deriving the fundamental equations, we prove the collapse and amplification mechanism, the two most important features of a consistent collapse model. Under reasonable simplifying assumptions, we constrain the strength $\xi$ of the complex metric fluctuations with available experimental data. We show that $\xi\geq 10^{-26}$ in order for the model to guarantee classicality of macro-objects, and at the same time $\xi \leq 10^{-20}$ in order not to contradict experimental evidence. As a comparison, in the recent discovery of gravitational waves in the frequency range 35 to 250 Hz,  the (real) metric fluctuations reach a peak of $\xi \sim 10^{-21}$.
\end{abstract}
\maketitle

\section{Introduction}

The possibility for quantum mechanics to be the limiting case of an
underlying nonlinear theory has been often considered in the literature~\cite{leggett1980macroscopic,drever1983quantum,Penrose1986PENQCI,clark1987macroscopic,PhysRevLett.62.485,adler2004quantum}.
A straightforward motivation is that linear models typically  are an
approximation of nonlinear ones~\cite{PhysRevLett.62.485}. A stronger motivation is that they
open the way to solving the quantum measurement problem~\cite{bell2004speakable}.
In this latter context, models of spontaneous wave function collapse~\cite{PhysRevD.34.470,ghirardi1990markov,bassi2003dynamical,RevModPhys.85.471} provide a consistent phenomenology describing the collapse
of the wave function during a measurement, via extra nonlinear and
stochastic terms added to the dynamics. Due to their intrinsic nonlinearity,
these models also offer a way out for some of the puzzles in quantum
gravity and cosmology~\cite{PhysRevD.91.124009,Modak2015,PhysRevD.94.045009}.

The common feature of all collapse models is a classical noise, coupled
nonlinearly to the quantum wave function. The typical collapse equation,
in the It\^o form, is: 
\begin{align}
d\psi_{t}=&\left[-\frac{i}{\hbar}\hat{H}_{0}dt+\sqrt{\lambda}\sum_j(\hat{A_j}-\langle \hat{A_j}\rangle_{t})dW_{j,t}\right. \nonumber \\
&\left.-\frac{\lambda}{2}\sum_j(\hat{A_j}-\langle \hat{A_j}\rangle_{t})^{2}dt\right]\psi_{t}, \label{eq:1}
\end{align}
where $\hat{H}_0$ is the standard quantum Hamiltonian, $\{\hat{A_j}\}_j$ is a set of self-adjoint
commuting operators, and $\langle \hat{A_j}\rangle_{t}=\langle\psi_{t}|\hat{A_j}|\psi_{t}\rangle$
and $W_{j,t}$ are a set of independent Wiener processes, which force the wave function to collapse
towards one of the common eigenstates of the operators $\hat{A_j}$~\cite{0305-4470-34-42-306}.
The positive coupling constant $\lambda$ sets the strength of the
collapse mechanism.

Eq.~\eqref{eq:1} should be considered as a phenomenological equation,
raising the question of why it takes that form. A justification comes
from the following argument first proposed by Adler~\cite{adler2004quantum}.
Consider the Hamiltonian: 
\begin{align}
\hat{H}=\hat{H}_{0}+i\hbar\sqrt{\lambda}\sum_j\hat{A_j}\,w_{j,t},\label{eq:2}
\end{align}
where $w_{j,t}=dW_{j,t}/dt$ is a set of independent white noises. It describes the coupling
of a quantum system with external classical noises, through the
operators $\hat{A_j} $. It is a reasonable phenomenological ansatz, except
for the fact that the second term is anti-Herimitian~\footnote{A possible hint for this anzatz can be found in the works by M. B. Mensky et al.~\cite{menskii1979quantum, mensky1979quantum, golubtsova1989quantum, mensky1993continuous}. In particular, in~\cite{golubtsova1989quantum} it is shown how the presence of a continuous measurement (whose effects are expected to be similar to those of a continuous collapse) can be described effectively by adding an anti-hermiant term to the Hamiltonian of the system.}. 
As a consequence,
the norm of $\psi_{t}$ is not conserved, jeopardizing the physical
meaning of the wave function. The obvious thing to do is to replace
$\psi_{t}$ with $\psi_{t}/\|\psi_{t}\|$, but this brings in a serious
problem: the resulting equation is nonlinear and also the stochastic
ensemble of states evolves nonlinearly, even in the average. This
leads to superluminal signaling~\cite{Gisin1989}. The problem can be avoided
if one adds extra terms in Eq.~\eqref{eq:2}, such that the master
equation for density matrix $\rho_{t}=\mathbb{E}[|\psi_{t}\rangle\langle\psi_{t}|]$
associated with the ensemble becomes linear (and of the Lindblad type~\cite{Lindblad1976,GKS1976,Breuerbook}). These new terms are precisely those, which lead to Eq.~\eqref{eq:1}.
Appendix A contains the  derivation of what outlined here.

In the sense explained here above, the requirements of norm conservation
and no-superluminal signaling added to Eq.~\eqref{eq:2}, give the
desired collapse equation. The hope is that a sensible nonlinear pre-quantum
theory, which leads to a dynamics for the wave function at the phenomenological
level, will naturally embody both requirements. The open issue now,
is how to justify $\hat{H}$ in~\eqref{eq:2}, in particular why the coupling
should be anti-Herimitian, and what is the suitable choice for the
operators $\hat{A_j}$, which select the basis along which the collapse occurs.
While there is no answer to the first question -- at least no more than the hope
that the pre-quantum theory will provide a natural answer -- one can
say more about the second question.

Quite often the literature suggests that the collapse is driven by
gravity~\cite{karolyhazy1966gravitation,penrose1986quantum,miller1990sixty,frenkel2002tentative,diosi1987universal,PhysRevA.40.1165,Diosi2007DP,Penrose1996gravity}. This is the only possibility one can have, to link the collapse to a known force, since all other forces as we know them
have been successfully quantized therefore, they cannot provide the
anti-Herimitian coupling needed for the non-linear collapse. But there
is a stronger motivation. The collapse scales with the mass/size of the
system \cite{PhysRevD.34.470,ghirardi1990markov}, and localizes the wave function in space. Then,
the natural candidate for the operators $\hat{A_j}$ is the \textit{local mass
density} $\hat{m}({\bf x})=\sum_{i}m_{i}\delta^{(3)}({\bf x}-\hat{{\bf x}}_{i})$,
coupled to a noise $w({\bf x},t)$ spread through space~\footnote{Here the discrete label ``$j$" is replaced by the continuous label $\ve{x}$ representing the space points. Accordingly the discrete sums over $j$ becomes an integral over $\ve{x}$.}: 
\begin{align}
\hat{H} = \hat{H}_0 + i\hbar\sqrt{\xi}\int d^{3}x\, \hat{m}({\bf x})w({\bf x},t).\label{eq:gf}
\end{align}
A random \textit{gravitational} field naturally provides such a coupling
(see Appendix B), which would contain an anti-Herimitian part if the field has an
imaginary component. In~\cite{Adler2015} arguments are presented, as to why
the metric could be classical and complex-valued. For example, complex-valued effective metrics  appear in modified gravity theories, when chiral deformations of general relativity are allowed~\cite{krasnov2015gr}. Following this idea we will explore the consequences of assuming a {\it complex non-white classical
noise coupled to the local mass density}. 

The paper is organized as follows. In Sec.~II we  derive, to the first meaningful perturbative order, the general collapse equation for the wave function, as well as the associated master equation, in the case of $N$ complex valued coloured random noises $h_i(t)$, each coupled to an operator $\hat{A}_i$. The literature so far considered only the case of real valued coloured noises~\cite{RevModPhys.85.471}. In Sec.~III we show the collapse mechanism. In Sec.~IV we consider specifically a noise field $w({\bf x},t)$ coupled to the local mass density $\hat{m}({\bf x})$ and discuss the amplification mechanism, one of the crucial properties of any collapse model. In Sec.~V we analyze the bounds on the spectrum of the noise, which are set by current experiments. We conclude the paper with a discussion of the results (Sec.~VI).

\section{Master and Collapse equations}

We have seen how the idea of a complex gravitational stochastic background inducing the collapse of the wave function leads to a collapse model where the noise is \textit{complex} valued and, in general,\textit{coloured}. Since this has not been discussed in the literature so far, in this section we derive the appropriate collapse equation and the master equation, following the same strategy as in Appendix A for a real valued white noise. 
The starting point is the following generalized Schr\"odinger
equation 
\begin{align}
{i\hbar}\partial_{t}\ket{\phi_{t}}=\left[\hat{H}_{0}+\xi\sum_{i=1}^{N}\hat{A}_{i}\st_{i}(t)+\hat{O}\right]\ket{\phi_{t}}\label{eq:sds}
\end{align}
where $\hat{A}_{i}$ are arbitrary self-adjoint operators and $\st_{i}(t)$
are $N$ complex Gaussian noises, with zero average and correlation
function: 
\begin{align}
\mathbb{E}_{\mathbb{Q}}[\st_{i}^{*}(t)\st_{j}(\tau)] & =D_{ij}(t,\tau),\nonumber \\
\mathbb{E}_{\mathbb{Q}}[\st_{i}(t)\st_{j}(\tau)] & =S_{ij}(t,\tau).  \label{correlation}
\end{align}
$D_{ij}(t,\tau)$ and $S_{ij}(t,\tau)$ are complex functions with
magnitudes of order $1$ and $\hat{O}$ is an operator yet
to be defined. The parameter $\xi$ sets the strength of the noise, which is assumed to be small. Following the scheme outlined in the introduction, we will determine $\hat{O}$ by the requirement of non-faster-than-light signaling.

Since the norm of $|\phi_{t}\rangle$ is not conserved, we consider the normalized state $\ket{\psi_{t}}=\ket{\phi_{t}}/\abs{\phi_{t}}$, which solves the equation 
\begin{align}\label{eq:Psit}
i\hbar\partial_{t}\ket{\psi_{t}} 
 & =\left(\hat{H}_{t}-\frac{1}{2}\braket{\hat{H}_{t}-\hat{H}_{t}^{\dagger}}_{t}\right)\ket{\psi_{t}}
\end{align}
with 
\begin{align}
\hat{H}_{t}=\hat{H}_{0}+\xi\sum_{i=1}^{N}\hat{A}_{i}\st_{i}(t)+\hat{O}.
\end{align}
As expected, the normalized vector evolves according to a nonlinear
stochastic dynamics. The stochastic ensemble of pure states $\rho_{t}^{\st}=\ket{\psi_{t}}\!\bra{\psi_{t}}$
obeys the following dynamics: 
\begin{align}\label{eq:stocmast}
i\hbar\partial_{t}\rho_{t}^{\st}  =&\Big[\hat{H}_{0}+\xi\sum_{i=1}^{N}\Big(\hat{A}_{i}\st_{i}(t)-i\braket{A_{i}}_{t}\st_{i}^{\I}(t)\nonumber\\
&+\hat{O}-\frac{1}{2}\braket{\hat{O}-\hat{O}^{\dagger}}_{t}\Big]\rho_{t}^{\st}-\text{H.c.}
\end{align}
Taking the expectation value to compute the dynamics for the density
matrix $\rho_{t}=\av{\rho_{t}^{\st}}$, one obtains in general a nonlinear evolution for the ensemble, which implies the possibility of faster-than-lightt signaling~\cite{Gisin1989}. This can be avoided with a proper choice of the operator $\hat{O}$. Contrary to the white-noise case,
identifying the correct form of $\hat{O}$ is very difficult (in general, impossible) since the dependence
of the right-hand-side of the above equation on the noise $h$ is
highly nontrivial. This means that one is not able to compute the stochastic average and without such knowledge, $\hat{O}$ cannot be determined. One way to circumvent the problem is to proceed perturbatively~\cite{1751-8121-40-50-012}.
We Taylor-expand $\rho_{t}^{\st}$ in terms of $\xi$: 
\begin{align}\label{eq:pertorddzero}
\rho_{t}^{\st}=\rho_{0,t}^{\st}+\xi\rho_{1,t}^{\st}+\xi^{2}\rho_{2,t}^{\st}+\mathcal{O}(\xi^{3})
\end{align}
where, for $t=0$, all terms except the first one are zero. We also expand $\hat{O}$ in powers of $\xi$~\footnote{In the expansion of $\hat{O}$ we do not include a zero order term i.e. one independent from $\xi$. This because $\hat{O}$ is introduced to add corrective terms due to the presence of the noise terms, and therefore is expected to be zero in the limit $\xi \rightarrow 0$.}: 
\begin{align}\label{eq:Oexp}
\hat{O}=\xi\hat{O}_{1}+\xi^{2}\hat{O}_{2}+\mathcal{O}(\xi^{3}).
\end{align}
Exploiting the perturbative series above, one can find a closed equation for the average density $\av{\rho_{t}^{h}}$ and obtain the explicit expression \eqref{Eq.Oseries} for each term of the series \eqref{eq:Oexp}, such that the average dynamic does not produce faster-than-lightt signaling (see Appendix C for the detailed calculation).
This fixes the dynamical equation for the averaged density matrix to be, up to the second order in $\xi^{2}$:
\begin{align}\label{mefinal}
&\partial_{t}{\rho}_{t}  =-\frac{i}{\hbar}\com{\hat{H}_{0}}{\rho_{t}}\nonumber\\
&-\frac{\xi^{2}}{\hbar^{2}}\Bigg[\sum_{i,j=1}^{N}\int_{0}^{t}\!\!d\tau\,D_{ij}^{\text{\tiny R}}(t,\tau)\com{\hat{A}_{i}}{\com{\hat{A}_{j}(\tau-t)}{\rho_{t}}}+\nonumber\\
 & i\sum_{i,j=1}^{N}\!\int_{0}^{t}\!\!\!d\tau D_{ij}^{\I}(t,\tau)\com{\hat{A}_{i}}{\acom{\hat{A}_{j}(\tau\!-\!t)}{\rho_{t}}}\Bigg]\!+\!\mathcal{O}(\xi^{3}),
\end{align}
 where the superscript R/I stands for the real/imaginary part~\footnote{Note that the correlator $S_{ij}(t,\tau)$ does not appear.}.
 Exploiting then Eq.~\eqref{Eq.Oseries} in Eq.~\eqref{eq:Psit}, the collapse equation for the wave function turns out to be, up to second order in  $\xi^{2}$:
\begin{widetext}
\begin{align}\label{eq:collapse}
i\hbar\partial_{t}\ket{\psi_{t}} & =\left[\hat{H}_{0}+{\xi}\left(\sum_{i=1}^{N}(\hat{A}_{i}\st_{i}(t)-i\braket{A_{i}}_{t}\st_{i}^{\I}(t)\right)\right.\nonumber \\
 & +\frac{i\xi^{2}}{\hbar}\sum_{i,j=1}^{N}\int_{0}^{t}d\tau\left(S_{ij}(t,\tau)-D_{ij}(t,\tau)\right)\hat{A}_{i}(\hat{A}_{j}(t-\tau)-\braket{A_{j}(t-\tau)}_{t})\nonumber \\
 & -\frac{i\xi^{2}}{\hbar}\sum_{i,j=1}^{N}\int_{0}^{t}d\tau(S_{ij}(t,\tau)-D_{ij}^{*}(t,\tau))\braket{A_{i}}_{t}\hat{A}_{j}(\tau-t)\nonumber \\
 & -\frac{i\xi^{2}}{2\hbar}\sum_{i,j=1}^{N}\int d\tau(S_{ij}(t,\tau)-D_{ij}(t,\tau))(\braket{A_{i}A_{j}(\tau-t)}_{t}-2\braket{A_{i}}_{t}\braket{A_{j}(\tau-t)}_{t})\nonumber \\
 & \left.-\frac{i\xi^{2}}{2\hbar}\sum_{i,j=1}^{N}\int d\tau(S_{ij}^{*}(t,\tau)-D_{ij}^{*}(t,\tau))(\braket{A_{j}(\tau-t)A_{i}}_{t}-2\braket{A_{i}}_{t}\braket{A_{j}(\tau-t)}_{t})\right]\ket{\psi_{t}}.
\end{align}
It is interesting to write down the Markovian limit which is obtained by imposing 
$D_{ij}(t,s)=\delta(t-s)\tilde{D}_{ij}(t)$ and $S_{ij}(t,s)=\delta(t-s)\tilde{S}_{ij}(t)$;  one ends up with the following stochastic Schr\"odinger equation in the Stratonovich form:
\begin{align}
i\hbar\partial_{t}\ket{\psi_{t}} & =\left[\hat{H}_{0}+{\xi}\left(\sum_{i=1}^{N}(\hat{A}_{i}\st_{i}(t)-i\braket{A_{i}}_{t}\st_{i}^{\I}(t)\right)\right.\nonumber \\
&+\frac{i\xi^{2}}{\hbar}\sum_{i,j=1}^{n}(\tilde{S}_{ij}(t)-\tilde{D}_{ij}(t))[(\hat{A}_{i}-\braket{A_{i}}_{t})(\hat{A}_{j}-\braket{A_{j}}_{t})+\frac{1}{2}(\braket{A_{i}A_{j}}_{t}+\braket{A_{j}A_{i}}_{t}-2\braket{A}_{i}\braket{A}_{j})]\nonumber\\
&\left.-\frac{i\xi^{2}}{\hbar}\sum_{i,j=1}^{N}(\tilde{S}_{ij}^{\I}(t)-\tilde{D}_{ij}^{\I}(t))\left(\braket{A_{i}A_{j}}_{t}-2\braket{A_{j}A_{i}}_{t} \right) +\frac{i2\xi}{\hbar}\sum_{i,j=1}^{n}\tilde{D}_{ij}^{\I}(t)\braket{A_{i}}_{t}\hat{A}_{j}\right]\ket{\psi_{t}}.
\end{align}
\end{widetext}
This equation is a generalization of Eq.~(7.43) in~\cite{bassi2003dynamical}. The first two lines correspond to Eq.~(7.43), with the replacement $\gamma \rightarrow \tilde{S}_{ij}(t)-\tilde{D}_{ij}(t)$, taking also into account that in our case the operators $A_i$ are not assumed to commute; the third line is associated with the complex part of the noise, while in~\cite{bassi2003dynamical} the noise was assumed to be real.

Equations.~\eqref{mefinal} and~\eqref{eq:collapse} are the main result of this section, and will be used in the rest of the work.

 In the next sections we will discuss the main consequences of Eqs.~\eqref{mefinal} and~\eqref{eq:collapse}: the collapse of the wave function, the presence, under suitable conditions, of an amplification mechanism, and some experimental predictions.

\section{Collapse of the wave function}

We now establish under which conditions the dynamics given by Eq.~\eqref{eq:collapse}, when $H_{0}=0$,
induce the collapse of the state vector $\ket{\psi}_{t}$ into one
of the eigenstates of $\hat{A}_{i}$, assuming that these operators commute with each other and therefore have a common set of eigenstates.
We will follow the procedure outlined in Sec.~IIa of~\cite{1751-8121-40-50-012}. We neglect the standard evolution since we are focusing only on the collapse process. This approximation, in general not true, is good  for macroscopic objects. In fact, given the amplification mechanism, which we will describe in the next section, the effect of the collapse increases with the mass of the system, becoming  dominant with respect to the standard evolution for large objects. 

We consider the stochastic average of the variance $V_{A}(t)=\braket{\hat{A}^{2}}_{t}-\braket{\hat{A}}_{t}^{2}$
of an operator $\hat{A}$ which commutes with all $\hat{A}_{i}$. 
 One may prove that, for any $n$: 
\begin{align}
\av{\braket{\hat{A}^{n}}_{t}}=\Tr[\rho_{t}\hat{A}^{n}]=\Tr[\rho_{0}\hat{A}^{n}]=\av{\braket{\hat{A}^{n}}_{0}}.\label{eq:An}
\end{align}
Then, exploiting the perturbative series in Eq.~\eqref{eq:pertordd}
and performing the stochastic average one can obtain: 
\begin{align}
 & \av{\braket{A}_{t}^{2}}= \av{\braket{A}_{0}} \nonumber \\
 & -\frac{2\xi^{2}}{\hbar^{2}}\sum_{i,j=1}^{N}\int_{0}^{t}\!\!\!d\tau\int_{0}^{\tau}\!\!\!ds\,\Bigl(S_{ij}^{\R}(\tau,s)-D_{ij}^{\R}(\tau,s)\Bigr)\times\nonumber\\
 &\braket{\braket{A}_{0}(A_{i}-\braket{A_{i}}_{0}}_{0} \braket{\braket{A}_{0}(A_{j}-\braket{A_{j}}_{0}}_{0}+\mathcal{O}(\xi^{3}).
\end{align}
Given the above result, one can now compute the stochastic average
of the variance $V_{A}(t)$, arriving at: 
\begin{align}\label{eq:variancev}
&\av{V_{A}(t)}=  V_{A}(0)-\frac{2\xi^{2}}{\hbar^{2}}\sum_{i,j=1}^{N}\int_{0}^{t}\!\!\!\!d\tau\,F_{ij}(\tau)\times\nonumber\\
&\Bigl(\braket{A\,A_{i}}_{0}-\braket{A}_{0}\braket{A_{i}}_{0}\Bigr)\Bigl(\braket{A\,A_{j})}_{0}-\braket{A}_{0}\braket{A_{j}}_{0}\Bigr)\nonumber\\&+\mathcal{O}(\xi^{3})
\end{align}
where:
\begin{align}
F_{ij}(\tau)=\int_{0}^{\tau}d\s\left(D_{ij}^{\R}(\tau,s)-S_{ij}^{\R}(\tau,s)\right).
\end{align}
According to~\footnote{See p.~9 in~\cite{1751-8121-40-50-012}.} the positivity
of $F(\ve{x},\ve{y},\tau)$ in the limit $t\to\infty$ is a sufficient
condition to guarantee the reduction properties of Eq.~\eqref{eq:collapse}.
In fact, whenever $F$ is non-negative, Eq.~\eqref{eq:variancev} implies
that, for large times $\braket{A\,A_{i}}_{\tau}-\braket{A}_{\tau}\braket{A_{i}}_{\tau}$
converges to 0 for any realization of the noise, with the only possible
exception of a subset of measure 0. In particular, when $\hat{A}$
is equal to $A_{i}$ we have
\begin{align}
\lim_{t\to\infty}\braket{A_{i}\,A_{i}}_{t}-\braket{A_{i}}_{t}\braket{A_{i}}_{t}=\lim_{t\to\infty}V_{A_{i}}(t)=0.
\end{align}
This means that any initial state converges asymptotically, with probability
1, to one of the eigenstates of the operator $\hat{A_{i}}$.

A related question is how fast the wave function collapses. The decoherence rate of the associated master Eq.~\eqref{mefinal} provides a good measure.
If we set $\hat{H}_{0}=0$, we immediately obtain the decoherence rate in the basis of the common eigenstates of the operators $\hat{A}_{i}$:
\begin{widetext}
\begin{align}\label{eq:ratedeco}
\rho_{t}(\alpha,\beta)&= \exp\left( -\frac{\xi^{2}}{\hbar^{2}}\sum_{i,j=1}^{N}\int_{0}^{t}d\tau \int_{0}^{\tau}ds D_{ij}^{\R}(\tau,s)(\alpha_{i}\alpha_{j}-\alpha_{i}\beta_{j}-\alpha_{j}\beta_{i}+\beta_{j}\beta_{i})\right.\nonumber\\
&+iD_{ij}^{\I}(\tau,s)(\alpha_{i}\alpha_{j}+\alpha_{i}\beta_{j}+\alpha_{j}\beta_{i}+\beta_{j}\beta_{i}) \Bigg)\rho_{0}(\alpha,\beta)=\nonumber\\
&=\exp\left(-\frac{\xi^{2}}{\hbar^{2}}\sum_{i,j=1}^{N}\int_{0}^{t}d\tau \int_{0}^{\tau} ds D_{ij}(\tau,s)(\alpha_{i}\alpha_{j}-\beta_{i}\beta_{j})-D_{i,j}^{*}(\tau,s)(\alpha_{i}\beta_{j}+\alpha_{j}\beta_{i}) \right)\rho_{0}(\alpha,\beta)
\end{align}
\end{widetext}
where $\rho_{t}(\alpha,\beta)= \bra{\alpha}\rho_{t}\ket{\beta}$ and $\ket{\alpha}$ $(\ket{\beta})$ is one element of the basis, \textit{i.e.} $\hat{A}_{i}\ket{\alpha}=\alpha_{i}\ket{\alpha}$.

It is worth studying the case where there is only one collapse operator and the correlation is real and delta correlated in time, \textit{i.e.}
\begin{align}\label{taudef}
D(\tau,s)= \tau_{0}\,\delta(\tau-s)
\end{align}
with $\tau_{0}$ a real parameter with the dimensions of a time. Then Eq.~\eqref{eq:ratedeco} reduces to:
\begin{align}
\rho_{t}(\alpha,\beta)= e^{-\frac{\xi^{2}\tau_{0}t}{\hbar^{2}}(\alpha-\beta)^{2}}\rho_{0}(\alpha,\beta)
\end{align}
where the decoherence rate is constant in time and is determined by $\tau_{0}\xi^{2}$.

\section{Master Equation for the center of mass and the amplification mechanism\label{sec:Master-Equation}}

After the collapse of the wave function, the next fundamental requirement
for a good collapse model is the  amplification
mechanism: the center of mass wave function of a composite system should collapse with a rate which increases with the size of the system. This is necessary in order for the  equation to preserve the quantum properties of microscopic systems and, at the same time, to guarantee the classical properties of macroscopic objects. 

Instead of considering the problem in full generality as done in the previous two sections, we focus our analysis to the case of interest here: the collapse noise coupled to the mass density operator $\hat{m}(\ve{x})$.
In this case Eq.~\eqref{mefinal} takes the form: 
\begin{align}
\begin{split}\partial_{t}{\rho}_{t}  =&-\frac{i}{\hbar}\com{\hat{H}_{0}}{\rho_{t}}\\
  -&\frac{\xi^{2}c^4}{\hbar^{2}}\int d\mathbf{x}\int d\mathbf{y}\int_{0}^{t}d\tau\,D^{\R}(\mathbf{x}-\mathbf{y},t-\tau) \\
 & \times \com{\hat{m}(\mathbf{x})}{\com{\hat{m}(\mathbf{y},\tau-t)}{\rho_{t}}}\\
  -&\frac{i\xi^{2}c^4}{\hbar^{2}}\int d\mathbf{x}\int d\mathbf{y}\int_{0}^{t}d\tau\,D^{\I}(\mathbf{x}-\mathbf{y},t-\tau) \\ 
& \times \com{\hat{m}(\mathbf{x})}{\acom{\hat{m}(\mathbf{y},\tau-t)}{\rho_{t}}},
\end{split}
\label{new1}
\end{align}
where $D^{\R}$ and $D^{\I}$ are the real and the imaginary parts
of the correlation function of the noise field~\footnote{Here again the discrete label $i$ is replaced by the continuous parameter $\ve{x}$.} $D(\ve{x},\ve{y};t,\tau)=\mathbb{E}[\st^{*}(\ve{x},t)\st(\ve{y},\tau)]$.
In writing the above equation, we assumed that the noise 
is statistically homogeneous over space and time: $D^{\R,\I}(\ve{x},\ve{y},t,\tau)=D^{\R,\I}(\ve{x}-\ve{y},t-\tau)$.
 We consider a system of $N$ pointlike particles. The mass density
function is: 
\begin{align}
\begin{split}
\hat{m}(\mathbf{x}) &=\sum_{i=1}^{N}m_{i}\delta(\mathbf{x}-\hat{\mathbf{x}}_{i})\\
&=\sum_{i=1}^{N}\frac{m_{i}}{(2\pi\hbar)^{3}}\int d\mathbf{Q}\,e^{\frac{i}{\hbar}\mathbf{Q}\cdot(\mathbf{x}-\hat{\mathbf{x}}_{i})}.\label{new2}
\end{split}
\end{align}
Substituting Eq.~(\ref{new2}) into Eq.~(\ref{new1})
and performing the integration over $\mathbf{x}$ and $\mathbf{y}$
we arrive at the expression:
\begin{align}
&\partial_{t}{\rho}_{t} =-\frac{i}{\hbar}\left[\hat{H}_{0},\rho_{t}\right]\nonumber\\
 & -\frac{\xi^{2}c^4}{\hbar^{2}}\sum_{i,j=1}^{N}\frac{m_{i}m_{j}}{(2\pi\hbar)^{3}}\int_{0}^{t}d\tau\int d\mathbf{Q}\,\tilde{D}^{\R}(\mathbf{Q},t-\tau)\nonumber\\
&\quad \times \left[e^{-\frac{i}{\hbar}\mathbf{Q}\cdot\hat{\mathbf{x}}_{i}},\left[e^{\frac{i}{\hbar}\hat{H}_{0}(\tau-t)}e^{\frac{i}{\hbar}\mathbf{Q}\cdot\hat{\mathbf{x}}_{j}}e^{-\frac{i}{\hbar}\hat{H}_{0}(\tau-t)},\rho_{t}\right]\right]\nonumber\\
 & -\frac{i\xi^{2}c^4}{\hbar^{2}}\sum_{i,j=1}^{N}\frac{m_{i}m_{j}}{(2\pi\hbar)^{3}}\int_{0}^{t}d\tau\int d\mathbf{Q}\,\tilde{D}^{\I}(\mathbf{Q},t-\tau)\nonumber\\
&\quad \times \left[e^{-\frac{i}{\hbar}\mathbf{Q}\cdot\hat{\mathbf{x}}_{i}}\!,\!\left\{ e^{\frac{i}{\hbar}\hat{H}_{0}(\tau-t)}e^{\frac{i}{\hbar}\mathbf{Q}\cdot\hat{\mathbf{x}}_{j}}e^{-\frac{i}{\hbar}\hat{H}_{0}(\tau-t)},\rho_{t}\right\} \right],
\label{new3}
\end{align}
where we defined:
\begin{align}
\tilde{D}^{\beta}(\mathbf{Q},t-\tau):=\int d\mathbf{r}\,D^{\beta}(\mathbf{r},t-\tau)e^{\frac{i}{\hbar}\mathbf{Q}\cdot\mathbf{r}}\label{new4}
\end{align}
with $\beta=\textrm{R},\,\textrm{I}$. 

We are interested in describing the dynamics of the center of mass
of the composite system. In particular, we have in mind the case of
a rigid body. We introduce the center of mass coordinates: 
\begin{align}
\hat{\ve{X}}=\sum_{i=1}^{N}\frac{m_{i}}{M}\hat{\ve{x}}_{i}\,, &  & \hat{\ve{P}}=\sum_{i=1}^{N}\hat{\ve{q}_{i}}\,,
\end{align}
and the relative coordinates 
\begin{eqnarray}
\left\{ \begin{array}{l}
\hat{\bm{r}}_{i}=\hat{\ve{x}_{i}}-\hat{\ve{X}}\quad\;\;\;\;\;\;\,\;\;\;\;\;  i\in(1,\dots,N-1)\,,\\
\\
{ \hat{\bm{r}}_{N}=-\sum_{i=1}^{N-1}\frac{m_{i}}{m_{N}}\hat{\bm{r}}_{i}\,,}\\
\\
{\hat{\ve{p}_{i}}=\hat{\ve{q}}_{i}-\frac{m_{i}}{M}\hat{\ve{P}}\quad\;\;\;\;\;\;\;}  i\in(1,\dots,N-1)\,,\\
\\
{\hat{\ve{p}}_{N}=-\sum_{i=1}^{N-1}\hat{\ve{p}_{i}}\,,}
\end{array}\right. 
\nonumber \\
\label{new5}
\end{eqnarray}
where $M=\sum_{i=1}^{N}m_{i}$ is the total mass of the system. The
operators $\hat{\bm{r}}_{N}$ and $\hat{\ve{p}}_{N}$ are not independent
(they are defined in terms of the other relative positions and momenta)
but it is convenient to keep them to make the notation simpler.
These new variables obey to the following commutation relations: 
\begin{align}
 & \com{\hat{\ve{X}}}{\hat{\ve{P}}}=i\hbar\;\;\;\;\;\;\;\;\com{\hat{\bm{r}}_{i}}{\hat{\ve{p}}_{j}}=i\hbar\left(\delta_{ij}-\frac{m_{i}}{M}\right)\nonumber \\
\nonumber \\
 & \com{\hat{\ve{X}}}{\hat{\bm{r}}_{i}}=\com{\hat{\ve{X}}}{\hat{\ve{p}}_{i}}=\com{\hat{\bm{r}}_{i}}{\hat{\bm{r}}_{j}}=\com{\hat{\bm{r}_{i}}}{\hat{\ve{P}}}=0
\end{align}
for $i,j\in(1,\dots,N-1)$. We introduce the center of mass density matrix as 
\[
\rho_{t}^{\textrm{\tiny CM}}:=\textrm{Tr}_{\textrm{\tiny REL}}\left(\rho_{t}\right)
\]
where $\textrm{Tr}_{\textrm{\tiny REL}}\left(\cdot\right)$
denotes the partial trace over the relative coordinates. We study
the effect of the partial trace on the operators of Eq.~(\ref{new3}). Assuming that $\hat{H}_{0}=\hat{H}_{0}^{\textrm{\tiny CM}}+\hat{H}_{0}^{\textrm{\tiny REL}}$,
the term in the first line simplifies as 
\begin{align*}
\textrm{Tr}_{\textrm{\tiny REL}}([\hat{H}_{0}^{\textrm{\tiny CM}}+\hat{H}_{0}^{\textrm{\tiny REL}},\rho_{t}])=[\hat{H}_{0}^{\textrm{\tiny CM}},\rho_{t}^{\textrm{\tiny CM}}].
\end{align*}
The double commutator in the third line can be expanded as the sum of
four terms: 
\begin{align}\label{new6}
 & \textrm{Tr}_{\textrm{\tiny REL}}\!\!\left(\!\left[e^{-\frac{i}{\hbar}\mathbf{Q}\cdot\hat{\mathbf{x}}_{i}},\!\!\left[e^{\frac{i}{\hbar}\hat{H}_{0}(\tau-t)}e^{\frac{i}{\hbar}\mathbf{Q}\cdot\hat{\mathbf{x}}_{j}}e^{-\frac{i}{\hbar}\hat{H}_{0}(\tau-t)},\!\rho_{t}\right]\!\right]\!\right)\nonumber\\
&=\textrm{Tr}_{\textrm{\tiny REL}}\left(e^{-\frac{i}{\hbar}\mathbf{Q}\cdot\hat{\mathbf{x}}_{i}}e^{\frac{i}{\hbar}\hat{H}_{0}(\tau-t)}e^{\frac{i}{\hbar}\mathbf{Q}\cdot\hat{\mathbf{x}}_{j}}e^{-\frac{i}{\hbar}\hat{H}_{0}(\tau-t)}\rho_{t}\right)\nonumber \\
 & -\textrm{Tr}_{\textrm{\tiny REL}}\left(e^{\frac{i}{\hbar}\hat{H}_{0}(\tau-t)}e^{\frac{i}{\hbar}\mathbf{Q}\cdot\hat{\mathbf{x}}_{j}}e^{-\frac{i}{\hbar}\hat{H}_{0}(\tau-t)}\rho_{t}e^{-\frac{i}{\hbar}\mathbf{Q}\cdot\hat{\mathbf{x}}_{i}}\right)\nonumber\\
&-\textrm{Tr}_{\textrm{\tiny REL}}\left(e^{-\frac{i}{\hbar}\mathbf{Q}\cdot\hat{\mathbf{x}}_{i}}\rho_{t}e^{\frac{i}{\hbar}\hat{H}_{0}(\tau-t)}e^{\frac{i}{\hbar}\mathbf{Q}\cdot\hat{\mathbf{x}}_{j}}e^{-\frac{i}{\hbar}\hat{H}_{0}(\tau-t)}\right)\nonumber \\
 & +\textrm{Tr}_{\textrm{\tiny REL}}\left(\rho_{t}e^{\frac{i}{\hbar}\hat{H}_{0}(\tau-t)}e^{\frac{i}{\hbar}\mathbf{Q}\cdot\hat{\mathbf{x}}_{j}}e^{-\frac{i}{\hbar}\hat{H}_{0}(\tau-t)}e^{-\frac{i}{\hbar}\mathbf{Q}\cdot\hat{\mathbf{x}}_{i}}\right)
\end{align}
We consider the first term on the right-hand side, as the calculations for the remaining terms are similar. Exploiting the commutativity of the relative and
center of mass degree of freedoms, we rewrite the exponential operators
in Eq.~(\ref{new6}) as 
\begin{align}
e^{-\frac{i}{\hbar}\mathbf{Q}\cdot\hat{\mathbf{x}}_{i}}&=e^{-\frac{i}{\hbar}\mathbf{Q}\cdot\hat{\bm{X}}}e^{-\frac{i}{\hbar}\mathbf{Q}\cdot\hat{\bm{r}}_{i}},\nonumber\\
 e^{\frac{i}{\hbar}\hat{H}_{0}(\tau-t)}&=e^{\frac{i}{\hbar}\hat{H}_{0}^{\textrm{\tiny CM}}(\tau-t)}e^{\frac{i}{\hbar}\hat{H}_{0}^{\textrm{\tiny REL}}(\tau-t)},
\end{align}
so that 
\begin{align}\label{new7}
 & \textrm{Tr}_{\textrm{\tiny REL}}\left(e^{-\frac{i}{\hbar}\mathbf{Q}\cdot\hat{\mathbf{x}}_{i}}e^{\frac{i}{\hbar}\hat{H}_{0}(\tau-t)}e^{\frac{i}{\hbar}\mathbf{Q}\cdot\hat{\mathbf{x}}_{j}}e^{-\frac{i}{\hbar}\hat{H}_{0}(\tau-t)}\rho_{t}\right)=
\nonumber \\
 & \textrm{Tr}_{\textrm{\tiny REL}}\left(\left[e^{\!-\frac{i}{\hbar}\mathbf{Q}\cdot\hat{\bm{r}}_{i}}e^{\frac{i}{\hbar}\hat{H}_{0}^{\textrm{\tiny REL}}(\tau-t)}e^{\frac{i}{\hbar}\mathbf{Q}\cdot\hat{\bm{r}}_{j}}e^{\!-\frac{i}{\hbar}\hat{H}_{0}^{\textrm{\tiny REL}}(\tau-t)}\right]\right.\nonumber\\
&\,\,\times\!\left.\left[e^{-\frac{i}{\hbar}\mathbf{Q}\cdot\hat{\bm{X}}}e^{\frac{i}{\hbar}\hat{H}_{0}^{\textrm{\tiny CM}}(\tau-t)}e^{\frac{i}{\hbar}\mathbf{Q}\cdot\hat{\bm{X}}}e^{-\frac{i}{\hbar}\hat{H}_{0}^{\textrm{\tiny CM}}(\tau-t)}\right]\rho_{t}\right).
\end{align}
We assume the motion of the relative coordinates to be a small fluctuation around the equilibrium positions $\bm{r}_{i0}$ within the solid (\textit{e.g.} in a crystalline structure),
\textit{i.e.} $\hat{\bm{r}}_{i}(t)=\bm{r}_{i0}+\Delta\hat{\bm{r}}_{i}(t)$, where the fluctuations $\Delta\hat{\bm{r}}_{i}(t)$ are negligible with respect to the spatial  correlation length of the noise within the time $t-\tau$ . Under this approximation, the square bracket in the second line of Eq.~\eqref{new7} becomes $e^{-\frac{i}{\hbar}\mathbf{Q}\cdot(\bm{r}_{i0}-\bm{r}_{j0})}$ and we obtain:
\begin{align*}
&\textrm{Tr}_{\textrm{\tiny REL}}\left(e^{-\frac{i}{\hbar}\mathbf{Q}\cdot\hat{\mathbf{x}}_{i}}e^{\frac{i}{\hbar}\hat{H}_{0}(\tau-t)}e^{\frac{i}{\hbar}\mathbf{Q}\cdot\hat{\mathbf{x}}_{j}}e^{-\frac{i}{\hbar}\hat{H}_{0}(\tau-t)}\rho_{t}\right)\simeq\\
&\simeq e^{-\frac{i}{\hbar}\mathbf{Q}\cdot(\bm{r}_{i0}-\bm{r}_{j0})}e^{-\frac{i}{\hbar}\mathbf{Q}\cdot\hat{\bm{X}}}e^{\frac{i}{\hbar}\hat{H}_{0}^{\textrm{CM}}(\tau-t)}e^{\frac{i}{\hbar}\mathbf{Q}\cdot\hat{\bm{X}}}\\
&\times e^{-\frac{i}{\hbar}\hat{H}_{0}^{\textrm{CM}}(\tau-t)}\rho_{t}^{\textrm{\tiny CM}},
\end{align*}
which depends on center of mass operators only. The other three terms on the right-hand side in Eq.~\eqref{new6} can be computed in the same way and therefore we get the overall result:
\begin{align*}
&\textrm{Tr}_{\textrm{\tiny REL}}\!\left(\!\left[e^{\!-\frac{i}{\hbar}\mathbf{Q}\cdot\hat{\mathbf{x}}_{i}}\!,\!\left[e^{\frac{i}{\hbar}\hat{H}_{0}(\tau-t)}e^{\frac{i}{\hbar}\mathbf{Q}\cdot\hat{\mathbf{x}}_{j}}e^{-\! \frac{i}{\hbar}\hat{H}_{0}(\tau-t)},\rho_{t}\right]\!\right]\!\right)\\
&=e^{\!-\frac{i}{\hbar}\mathbf{Q}\cdot(\bm{r}_{i0}-\bm{r}_{j0})}\times\\
&\left[e^{\!-\frac{i}{\hbar}\mathbf{Q}\cdot\hat{\bm{X}}}\!,\!\left[e^{\frac{i}{\hbar}\hat{H}_{0}^{\textrm{CM}}(\tau-t)}e^{\frac{i}{\hbar}\mathbf{Q}\cdot\hat{\bm{X}}}e^{-\frac{i}{\hbar}\hat{H}_{0}^{\textrm{CM}}(\tau-t)},\rho_{t}^{\textrm{\tiny CM}}\right]\right].
\end{align*}
Similarly, for the operators in the fifth line of Eq.~(\ref{new3})
we obtain: 
\begin{align*}
&\textrm{Tr}_{\textrm{\tiny REL}}\!\left(\!\left[e^{-\!\frac{i}{\hbar}\mathbf{Q}\cdot\hat{\mathbf{x}}_{i}}\!,\!\left\{ \!e^{\frac{i}{\hbar}\hat{H}_{0}(\!\tau-t)}e^{\frac{i}{\hbar}\mathbf{Q}\cdot\hat{\mathbf{x}}_{j}}e^{-\!\frac{i}{\hbar}\hat{H}_{0}(\tau-t)}\!,\rho_{t}\right\}\! \right]\!\right)\\
&=e^{-\frac{i}{\hbar}\mathbf{Q}\cdot(\bm{r}_{i0}-\bm{r}_{j0})}\times\\
&\left[e^{-\frac{i}{\hbar}\mathbf{Q}\cdot\hat{\bm{X}}}\!,\!\left\{ e^{\frac{i}{\hbar}\hat{H}_{0}^{\textrm{CM}}(\tau-t)}e^{\frac{i}{\hbar}\mathbf{Q}\cdot\hat{\bm{X}}}e^{-\!\frac{i}{\hbar}\hat{H}_{0}^{\textrm{CM}}(\tau-t)}\!,\rho_{t}^{\textrm{\tiny CM}}\right\} \right]\!.
\end{align*}
Combining the previous results, we arrive at the following master equation
for the center of mass 
\begin{align}
& \partial_{t}{\rho}_{t}^{\text{\tiny CM}}  =-\frac{i}{\hbar}\left[\hat{H}_{0}^{\textrm{\tiny CM}},\rho_{t}^{\text{\tiny CM}}\right] \nonumber\\
 & -\frac{\xi^{2}c^4}{\hbar^{2}}\frac{1}{(2\pi\hbar)^{3}}\int_{0}^{t}\!d\tau\!\int\! d\mathbf{Q}\,\tilde{D}^{\R}(\mathbf{Q},t-\tau)\,A(\mathbf{Q})\times\nonumber\\
&\left[e^{-\frac{i}{\hbar}\mathbf{Q}\cdot\hat{\bm{X}}}\!,\!\left[e^{\frac{i}{\hbar}\hat{H}_{0}^{\textrm{\tiny CM}}(\tau-t)}e^{\frac{i}{\hbar}\mathbf{Q}\cdot\hat{\bm{X}}}e^{-\frac{i}{\hbar}\hat{H}_{0}^{\textrm{\tiny CM}}(\tau-t)},\rho_{t}^{\textrm{\tiny CM}}\right]\right]\nonumber\\
&-\!\frac{i\xi^{2}c^4}{\hbar^{2}}\frac{1}{(2\pi\hbar)^{3}}\int_{0}^{t}\!d\tau\!\int \!d\mathbf{Q}\,\tilde{D}^{\I}(\mathbf{Q},t-\tau)\,A(\mathbf{Q})\times\nonumber\\
&\left[e^{-\frac{i}{\hbar}\mathbf{Q}\cdot\hat{\bm{X}}}\!,\left\{ e^{\frac{i}{\hbar}\hat{H}_{0}^{\textrm{\tiny CM}}(\tau-t)}e^{\frac{i}{\hbar}\mathbf{Q}\cdot\hat{\bm{X}}}e^{-\frac{i}{\hbar}\hat{H}_{0}^{\textrm{\tiny CM}}(\tau-t)},\rho_{t}^{\textrm{\tiny CM}}\right\}\! \right] 
\label{new8}
\end{align}
with: 
\begin{align}
A(\mathbf{Q}):=\sum_{i,j=1}^{N}m_{i}m_{j}e^{-\frac{i}{\hbar}\mathbf{Q}\cdot(\bm{r}_{i0}-\bm{r}_{j0})}=|\rho(\ve{Q}/\hbar)|^{2},\label{new9}
\end{align}
where
\begin{align}
{\rho}(\ve{k}):=\int d\ve{x}\rho(\ve{x})e^{-i\ve{k}\ve{x}}
\end{align}
is the Fourier transform of the classical mass density distribution $\rho(\ve{x}):=\sum_{i=1}^{N} m_i \delta(\ve{x}-\bm{r}_{i}^{cl})$.

The master equation~\eqref{new8} for the center of mass wave function has the same structure
as the single particle master equation, with the addition of the amplifying
factor $A(\mathbf{Q})$, which keeps track of the fact that we are dealing with a composite object, not a pointlike particle.

Typically, the noise correlators $D^{\R}(\mathbf{r},t-\tau)$ and $D^{\I}(\mathbf{r},t-\tau)$ are expected to have spatial cutoffs (the noise correlation length), respectively $r_{C}^{\R}$ and $r_{C}^{\I}$. As for the case of the  continuous spontaneous localization (CSL) model~\cite{ghirardi1990markov}, it is interesting to study the behavior of the amplification factor in two limiting cases (for a more detailed proof of what follows, see~\cite{torovs2016bounds2}): 

1.\!\!\!\!\!\!\!\! When the particles are at distances smaller than the noise correlation lengths
$r_{C}^{\R},\,r_{C}^{\I}$, they contribute \textit{coherently}, giving a factor $\propto\left(\sum_{i}m_{i}\right)^{2}$; 

2. When the particles are at distances larger than the noise correlation lengths
$r_{C}^{\R},\,r_{C}^{\I}$, they contribute \textit{incoherently} giving a factor
$\propto\sum_{i}m_{i}^{2}$. 

Because of these two properties, a reasonable estimate of the amplification factor in Eq.~(\ref{new9}), is provided by Adler's formula~\cite{adler2007lower,torovs2016bounds2}:
\begin{align}\label{new10}
A^{\beta}=A^{\beta}(r_{C}^{\beta})=N^{\beta}(n^{\beta}m_{0})^{2}\;\;\;\;\textrm{with}\;\;\;\;\beta=\textrm{R},\,\textrm{I},
\end{align}
where $A^{\beta}$ refers to $A$ in the second line of Eq.~\eqref{new8} for $\beta= \text{R}$ and to $A$ in the fourth line for $\beta= \text{I}$; $n^{\beta}$ is the number of nucleons of mass $m_{0}$ inside a sphere of radius $r_{C}^{\beta}$, while $N^{\beta}$ denotes the number of such spheres necessary for covering the entire object.

\section{Experimental bounds on the gravitational noise spectrum}

Discussing the experimental constraints on the noise correlator in its full generality is too difficult. We will limit the discussion to a restricted class of Gaussian correlations functions, in such a way that the collapse dynamics is controlled by only two parameters (for a class of correlation function that leads to a HPZ type master equation, see Appendix~\ref{HPZ}).

Specifically, we consider the Markovian limit by imposing
\begin{alignat}{1}
\tilde{D}^{\R}(\mathbf{Q},s)	\approx\tilde{D}^{\R}(\mathbf{Q})\,\tau_0\,\delta(s)\label{Markovian_limit}
\end{alignat}
with $[\tau_0]=[T]$ (see Eq.~\eqref{taudef}). From the definition of $D_{ij}(t,\tau)$  in Eq.~(\ref{correlation}), using the definition of the Fourier transform and Eq.~\eqref{Markovian_limit}, it is straightforward to show that $\tilde{D}^{\I}(\mathbf{Q})=0$. In addition, to make contact with existing phenomenology for the CSL model~\cite{ghirardi1990markov}, we assume that $\tilde{D}^{\R}(\mathbf{Q})$ has the following form:
\begin{align}
\tilde{D}^{\R}(\mathbf{Q})=r_C^3\,\text{exp}(-r_C^2\mathbf{Q}^{2}/\hbar^{2}),
\end{align}
where $[r_C]=[L]$. With these assumptions, after some algebra, Eq.~(\ref{new8}) reduces to
\begin{widetext}
\begin{align}\label{almostcslme}
&\partial_{t}{\rho}_{t}^{\text{\tiny CM}}=-\frac{i}{\hbar}\left[\hat{H}_{0}^{\textrm{\tiny CM}},\rho_{t}^{\text{\tiny CM}}\right]-\frac{\xi^{2}c^{4}r_{C}^{3}\tau_{0}}{(2\pi\hbar)^{3}2\hbar^{2}}\int d\mathbf{Q}\,A(\mathbf{Q})\text{exp}(-r_{C}^{2}Q^{2}/\hbar^{2})\left[e^{-\frac{i}{\hbar}\mathbf{Q}\cdot\hat{\bm{X}}},\left[e^{\frac{i}{\hbar}\mathbf{Q}\cdot\hat{\bm{X}}},\rho_{t}^{\textrm{\tiny CM}}\right]\right].
\end{align}
This equation should be compared with the CSL master equation \cite{ghirardi1990markov}:
\begin{align}\label{cslme}
\partial_{t}{\rho}_{t}^{\text{\tiny CM}}=-\frac{i}{\hbar}\left[\hat{H}_{0}^{\textrm{\tiny CM}},\rho_{t}^{\text{\tiny CM}}\right]-\frac{\lambda(4\pi r_C^2)^{3/2}}{(2\pi\hbar)^{3}}\int d\mathbf{Q}\,\frac{A(\mathbf{Q})}{m_0^2}\,\text{exp}(-r_C^2\mathbf{Q}^{2}/\hbar^{2})\,\left[e^{-\frac{i}{\hbar}\mathbf{Q}\cdot\hat{\bm{X}}},\left[e^{\frac{i}{\hbar}\mathbf{Q}\cdot\hat{\bm{X}}},\rho_{t}^{\textrm{\tiny CM}}\right]\right]
\end{align}   
\end{widetext}
In particular, Eq.~(\ref{almostcslme}) reduces to the CSL master equation given in Eq.~(\ref{cslme}) by setting:
\begin{align}
\xi=\frac{4\hbar\pi^{3/4}}{m_{0}c^{2}}\sqrt{\frac{\lambda}{\tau_{0}}}
\end{align}

\begin{figure}[h!]
\includegraphics[width=0.45\textwidth]{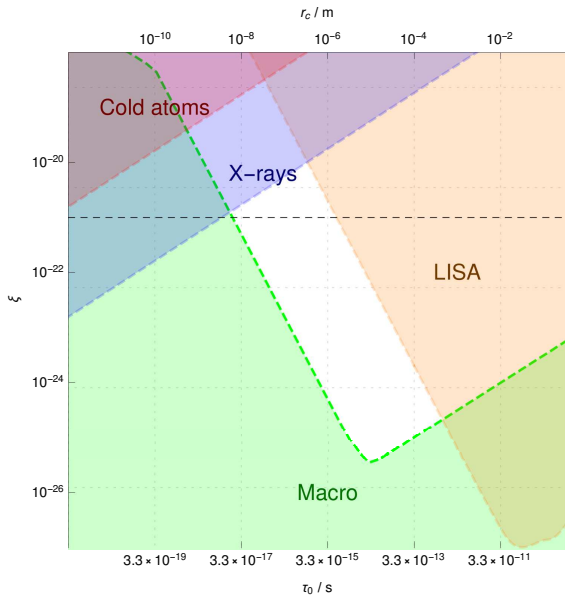}
\caption{$(\xi,r_C)$ or equivalently $(\xi,\tau_0)$ parameter diagram of the gravity-induced collapse model in the Markovian regime given by Eq.~\eqref{almostcslme}.  The white area is the allowed region. The other shaded regions are excluded: the orange shaded region ({\it LISA})  from the data analysis of LISA Pathfinder~\cite{carlesso}, the blue shaded region ({\it X-rays}) from data analysis of X-rays measurements~\cite{Curceanubeatrix},  the purple shaded region from the data analysis with cold atom experiments ({\it Cold atoms}) \cite{Bilardello2016764}. The green shaded region ({\it Macro}) is obtained by requiring that the collapse is strong enough to localize macroscopic objects~\cite{torovs2016bounds1,torovs2016bounds2}. As a reference, the horizontal dashed line is the magnitude of the real gravitational wave recently discovered by LIGO.
}
\centering
\label{Fig_parameters}
\end{figure}

To simplify the discussion, we assume the time cutoff to be related to the space cutoff via $\tau_0=r_C/c$. We can now set bounds on $(\xi,r_C)$ (or equivalently on $(\xi,\tau_0)$) by using the bounds already set for the CSL parameters $(\lambda,r_C)$. We have summarized the most recent bounds in Fig.~\ref{Fig_parameters}.

The primary feature of any good collapse model is to to suppress macroscopic linear superpositions.
By choosing an appropriate macroscopicity or classicality scale, one can estimate the minimal strength the collapse should have. Specifically, the lower bound ({\it Macro}) in Fig. 1 is obtained by requiring that an object of size $0.01$ mm is localized within $10$ ms~~\cite{torovs2016bounds1,torovs2016bounds2}. This means that more or less the smallest object visible to the naked eye is localized within the perception time of a human observer. Needless to say, this bound can change by several orders of magnitude depending on the  chosen criteria of classicality.

The coupling with the noise field not only suppress macroscopic superposition but, as a side effect, also makes particles constantly jiggle, and this random motion can be  tested by non-interferometric experiments~\cite{bahrami2014proposal, nimmrichter2014optomechanical, diosi2015testing}. Here we consider some of the most relevant such experiments, which set rather stringent bounds on the collapse parameters. The results are summarized in Fig.~1.

Consider first a charged particle: It is expected that the random jitters (accelerations) make it emit photons. The absence of this extra radiation, as it can be extracted by analyzing the spectrum of emission from Germanium measured over long times~\cite{Curceanubeatrix}  can then be used to obtain very good bounds bounds ({\it X-rays}). 

Another interesting bound on the size of the Brownian motion induced by the collapse mechanism comes from accurate monitoring of the motion of relatively large masses, as it is the case of  the LISA pathfinder experiment~\cite{carlesso}. These bounds have been obtained from the bounds on the parameters of the Markovian CSL model. However, for such a large object, we expect that the dynamics do not change significantly when we consider a colored noise: the relevant time scale of evolution is much longer than the considered values of the noise correlation time.

The last bound we consider is derived by studying the evolution of a gas of cold atoms. The collapse induced jiggling makes the gas expand more than what is predicted by standard quantum mechanics, and this difference becomes appreciable if the gas is initially at very low temperature. The absence of any appreciable difference~\cite{Bilardello2016764} gives the bound denoted by  "{\it Cold atoms}" in Fig.~1.  Although this bound is less strong than the one obtained from the X-ray experiment, it is the only one which has been shown to also persist for a non-Markovian noise field. 

We leave a more refined analysis of the other bounds, in the regime where non-Markovian are expected to become important, for future research.

We compare these results with the recent discovery of gravitational waves~\cite{abbott2016observation}, observed in frequency range from 35 to 250 Hz and with a peak strain of $1.0 \times 10^{-21}$. Clearly, gravitational waves are real, while here the claim is that the collapse is caused by complex fluctuations of the metric. Also, gravitational waves typically have longer wavelengths, while here the relevant part of the  spectrum  is at high frequencies (Fig.~\ref{Fig_parameters}). However, it is interesting to see that in order to have an efficient wave function collapse, the complex fluctuations need not be very strong. They can well be several orders of magnitude weaker than the real gravitational waves recently discovered.
In turn, this could explain why these complex fluctuations, if really existing, have not yet been discovered.

\section{Discussion and Conclusions}

Gravity-related models of spontaneous wave function collapse are not new in the literature. We mention two of them. The Di\'{o}si-Penrose (DP) model \cite{diosi1987universal,PhysRevA.40.1165,Diosi2007DP,Penrose1996gravity} has the same structure as the model considered here, with two important differences: (i) the noise is real and white in time and (ii) the spatial correlation function is proportional to $G/|\ve{x}-\ve{y}|$.
Although the model is certainly appealing in many ways, we see no reason why the noise correlator should have such a special form. Typically noises have rather complicated correlation functions, which have little or no connection to the form of the interaction.

The Schr\"odinger-Newton equation~\cite{diosi1984gravitation,Penrose1996gravity,carlip2008quantum} descends from semiclassical gravity~\cite{semiclass1,rosenfeld1963quantization} and contains a gravitational self-interaction term, which tends to suppresses superpositions in space. However, as discussed in~\cite{bahrami2014schrodinger}, this equation is not of the collapse-model type, in particular, it is not capable of predicting the collapse of the wave function in space with the correct quantum probabilities.


In this paper we have investigated a novel proposal, where the collapse
mechanism is driven by a complex fluctuating metric, as first suggested by Adler~\cite{Adler2015}. The correlation function should have a non negligible contribution also from relatively high frequency components ($\sim 10^{15}$ Hz), contrary to the current search for gravitational waves, which is focuses on much lower frequencies.

By imposing the condition of no superluminal signaling (perturbatively up to the second order in the coupling constant $\xi$, which sets the magnitude of the gravitational noise), we derived the structure of the equation describing the evolution of the state vector [Eq.~\eqref{eq:collapse}]. We then proved that this equation defines a good collapse dynamics: it collapses the state vector to the eigenstates of the preferred basis (in our case, the position basis) and it has an amplification mechanism which guarantees that, even for small $\xi$, collapse effects become relevant for macroscopic objects. 

In the last section we discussed experimental bounds on the parameters of the model. Interestingly enough the magnitude of the complex fluctuations needed for the collapse to be compatible with experimental data, and to guarantee the localization of macroscopic objects, can be orders of magnitude smaller than the recently discovered gravitational waves. Very weak fluctuations suffice to justify classicality as predicted by collapse models.

\acknowledgements
We thank S.L. Adler for several useful comments. The authors acknowledge financial support from the University of Trieste (FRA 2016) and INFN. G.G. acknowledges financial support  from ICTP Trieste.


\appendix

\section{Justification of the collapse equation}

We present the procedure outlined in the introduction, to justify the collapse equation. Here, to keep the notation simple, we focus on the case with only one operator $\hat{A}$ and one noise $w_t$. The generalization to the model described in Eq.~(\ref{eq:2}) can be trivially done, since the noises are independent.

Let us consider the Hamiltonian $\hat{H}=\hat{H}_0+i\hbar(\sqrt{\lambda}\hat{A}w_{t}+ \hat{O})$
and, in the It\^{o} language, the stochastic differential equation: 
\begin{align}\label{eq:a3}
d\phi_{t}=\left[-i\hat{H}_0dt+\sqrt{\lambda}\hat{A}dW_{t}+\hat{O}\right]\phi_{t};
\end{align}
throughout this section, we set $\hbar=1$. We will fix the form of
$\hat{O}$ by requiring no superluminal signaling. 

The norm of $\phi_{t}$ is not conserved. In order to write the equation for the normalized vector
$\psi_{t}=\phi_{t}/\|\phi_{t}\|$, let us consider the process $N_{t}=\langle\phi_{t}|\phi_{t}\rangle$.
Using It\^{o} rules ($dN_{t}=\langle d\phi_{t}|\phi_{t}\rangle+\langle\phi_{t}|d\phi_{t}\rangle+\langle d\phi_{t}|d\phi_{t}\rangle$)
one proves that it satisfies the stochastic differential equation:
\begin{align}
&dN_{t}=\nonumber\\
&\left[2\sqrt{\lambda}\langle \hat{A}\rangle_{t}dW_{t}+\lambda\langle \hat{A}^{2}\rangle_{t}dt+\langle(\hat{O}^{\dagger}+\hat{O})\rangle_{t}dt\right]N_{t},
\end{align}
where we have defined $\langle \hat{A}\rangle_{t}=\langle\phi_{t}|\hat{A}|\phi_{t}\rangle/\|\phi_{t}\|^{2}=\langle\psi_{t}|\hat{A}|\psi_{t}\rangle$,
and similarly for all other operators. From this, one can derive the
equation for $N_{t}^{-1/2}$: 
\begin{align}
&dN_{t}^{-1/2}=\left[-\sqrt{\lambda}\langle \hat{A}\rangle_{t}dW_{t}\right.\nonumber\\
&\left.\!+\!\left(\frac{3}{2}\lambda\langle \hat{A}\rangle_{t}^{2}\!-\!\frac{1}{2}\lambda\langle \hat{A}^{2}\rangle_{t}\!-\!\frac{1}{2}\langle(\hat{O}^{\dagger}+\hat{O})\rangle_{t}\!\!\right)\!dt\right]N_{t}^{-1/2},
\end{align}
and next the equation for $\psi_{t}=\phi_{t}N_{t}^{-1/2}$: 
\begin{align}
d\psi_{t} =&  \left[-i\hat{H}_{0}dt+\sqrt{\lambda}(\hat{A}-\langle \hat{A}\rangle_{t})dW_{t}\right.\nonumber\\
&\left.+\lambda\left(\frac{3}{2}\langle \hat{A}\rangle_{t}^{2}-\frac{1}{2}\langle \hat{A}^{2}\rangle_{t}-\hat{A}\langle \hat{A}\rangle_{t}\right)dt\right. \nonumber \\
 &+\left.\left(\hat{O}-\frac{1}{2}\langle(\hat{O}^{\dagger}+\hat{O})\rangle_{t}\right)dt\right]\psi_{t}.
\end{align}
As we can see, the normalized vector evolves according to a nonlinear
stochastic dynamics. The stochastic ensemble of pure states $\rho_{t}^{W}=|\psi_{t}\rangle\langle\psi_{t}|$
obeys the following dynamics: 
\begin{widetext}
\begin{eqnarray}
d\rho_{t}^{W} & = & -i[H,\rho_{t}^{W}]+\lambda\left(4\langle \hat{A}\rangle_{t}^{2}\rho_{t}^{W}-\langle \hat{A}^{2}\rangle_{t}\rho_{t}^{W}-2\hat{A}\langle \hat{A}\rangle_{t}\rho_{t}^{W}-2\rho_{t}^{W}\hat{A}\langle \hat{A}\rangle_{t}-\hat{A}\rho_{t}^{W}\hat{A}\right)dt\nonumber \\
 &  & +\left(\hat{O}^{\dagger}\rho_{t}^{W}+\rho_{t}^{W}\hat{O}-\langle(\hat{O}^{\dagger}+\hat{O})\rangle_{t}\rho_{t}^{W}\right)dt+(\text{extra terms})\,dW_{t}.
\end{eqnarray}
\end{widetext}
When taking the expectation value to compute the dynamics for the
density matrix $\rho_{t}=\mathbb{E}[\rho_{t}^{W}]$, the ``extra terms''
average to 0, while the remaining terms generate a nonlinear evolution
for the ensemble. This can be avoided by choosing $O=-(\lambda/2)\hat{A}^{2}+2\lambda(\hat{A}-\langle \hat{A}\rangle_{t})\langle \hat{A}\rangle_{t}$,
in which case all nonlinear terms cancel, and the equation for $\rho_{t}$
becomes of the Lindblad type: 
\begin{align}
\frac{d}{dt}\rho_{t}=-i[\hat{H}_{0},\rho_{t}]-\frac{\lambda}{2}[\hat{A},[\hat{A},\rho_{t}]];
\end{align}
in turn, Eq.~\eqref{eq:a3} reduces to Eq.~\eqref{eq:1}. This completes
the argument.

\section{Non-relativistic coupling between a gravitational background and
the local mass density}

The action of a matter field in curved space is described by: 
\begin{align}
S= & \int d^{4}x\sqrt{-g}\,\mathcal{L}_{m}\label{eq:ma}
\end{align}
where $\mathcal{L}_{m}$ is the matter Lagrangian, $g_{\mu\nu}$ is
the metric tensor and $\sqrt{-g}=\sqrt{-\det[g_{\mu\nu}]}$. We consider
a perturbation $\st_{\mu\nu}$ around the flat metric $\eta_{\mu\nu}$,
and we Taylor expand the action around it: 
\begin{align}
S=&\int d^{4}x\Big[\mathcal{L}_{m}^{(0)}+\nonumber\\
&\hspace{0.4cm}\left.\left.\frac{1}{\sqrt{-\eta}}\frac{\partial(\sqrt{-g}\,\mathcal{L})}{\partial g_{\mu\nu}}\right|_{\eta_{\mu\nu}}\st_{\mu\nu}\right]+\mathcal{O}(\st^{\mu\nu}h^{\delta\sigma});\label{eq:s1o}
\end{align}
the apex $(0)$ denotes the quantities in the flat space-time $\eta_{\mu\nu}$.
The stress energy tensor associated to the Lagrangian $\mathcal{L}_{m}$
is defined as follows~\cite{carroll2004spacetime}: 
\begin{align}
T^{\mu\nu}=\frac{-2}{\sqrt{-g}}\frac{\partial(\sqrt{-g}\,\mathcal{L}_{m})}{\partial g_{\mu\nu}},
\end{align}
and Eq.~\eqref{eq:s1o} can be rewritten in the form 
\begin{align}
S=\int d^{4}x\left[\mathcal{L}_{m}^{(0)}-\frac{1}{2}h^{\mu\nu}T_{\mu\nu}^{(0)}\right],\label{eq:interact_action}
\end{align}
where from now on we neglect higher order terms. In the weak field
limit, gravity couples to matter through the stress energy tensor.

We now derive the non relativistic limit, for a Klein-Gordon Lagrangian:
\begin{align}
\mathcal{L}_{m}^{(0)}=\frac{-\hbar^{2}}{2m}\left(\eta^{\mu\nu}\partial_{\mu}{\psi}^{*}\partial_{\nu}\psi-\frac{(mc)^{2}}{\hbar^{2}}{\psi}^{*}\psi\right).\label{eq:kg}
\end{align}
The interacting Lagrangian becomes: 
\begin{align}
\mathcal{L}_{int}^{(0)}&=-\frac{1}{2}\st^{\mu\nu}T_{\mu\nu}^{(0)}\nonumber\\
&=\frac{\hbar}{2m}\!\left(\!\partial_{\mu}\psi^{*}\partial_{\nu}\psi(\st^{\mu\nu}\!\!-\!\st_{\rho}^{\rho}\eta^{\mu\nu})\!-\!\st_{\rho}^{\rho}\frac{mc^{2}}{2\hbar^{2}}\psi^{*}\psi\!\right)\label{eq:kgi}
\end{align}
The non-relativistic limit can be obtained by rewriting the relativistic
wave function as follows: 
\begin{align}
\psi(x)=e^{\frac{i}{\hbar}mcx_{0}}\varphi(x),\label{eq:non_rel}
\end{align}
and assuming that the following relation holds: 
\begin{align}
\left|\frac{mc}{\hbar}\varphi\right|\gg|i\partial_{\mu}\varphi|,\label{eq:disprel}
\end{align}
meaning that the rest energy associated to the field $\varphi$ is
much bigger than the momentum energy. Inserting Eq.~\eqref{eq:non_rel}
into Eqs.~\eqref{eq:kg} and~\eqref{eq:kgi} one obtains 
\begin{align}
\mathcal{L}_{m}^{(0)} & =-\frac{\hbar}{2m}\Big[\partial_{0}\varphi^{*}\partial_{0}\varphi\nonumber\\
&+i\frac{mc}{\hbar}(\varphi^{*}\partial_{0}\varphi-(\partial_{0}\varphi^{*})\varphi)+\partial_{i}\varphi^{*}\partial_{i}\varphi\Big]
\end{align}
and 
\begin{align}
\mathcal{L}_{int}^{(0)}&=-\frac{1}{2}\st^{\mu\nu}T_{\mu\nu}^{(0)}\nonumber\\
 & =-\frac{\hbar^{2}}{2m}\left[\st^{00}\left(\partial_{0}-i\frac{mc}{\hbar}\right)\varphi^{*}\left(\partial_{0}+i\frac{mc}{\hbar}\right)\varphi\right.\nonumber\\
 & -\!\st^{0i}\!\left[\left(\!\partial_{0}\!-\!i\frac{mc}{\hbar}\right)\varphi^{*}\partial_{i}\varphi+\partial_{i}\varphi^{*}\!\!\left(\partial_{0}\!+\! i\frac{mc}{\hbar}\right)\!\varphi\right]\nonumber \\
 &+(\st^{ij}-\st_{\rho}^{\rho}\eta^{ij})\partial_{i}\varphi^{*}\partial_{i} \varphi \Big].
\end{align}
Under the assumption in~\eqref{eq:disprel}, we arrive at the symmetrized
free Schr\"odinger Lagrangian ($x_{0}=ct$): 
\begin{align}
\mathcal{L}_{m}^{(0)}\simeq\frac{i\hbar}{2}(\varphi^{*}\partial_{t}\varphi-\partial_{t}(\varphi^{*})\varphi)+\frac{\hbar^{2}}{2m}\partial_{i}\varphi^{*}\partial^{i}\varphi
\end{align}
and the interaction Lagrangian 
\begin{align}
\mathcal{L}_{int}^{(0)}=-\frac{mc^{2}}{2}\st^{00}\varphi^{*}\varphi.
\end{align}

The conjugate momenta associated with the total Lagrangian $\mathcal{L}=\mathcal{L}_{m}^{(0)}+\mathcal{L}_{int}^{(0)}$
are 
\begin{align}
\pi&=\frac{\partial\mathcal{L}}{\partial(\partial_{t}\varphi)}=\frac{i\hbar}{2}\varphi^{*},\nonumber \\
\pi^{*}&=\frac{\partial\mathcal{L}}{\partial(\partial_{t}\varphi^{*})}=-\frac{i\hbar}{2}\varphi,
\end{align}
and the Hamiltonian density is: 
\begin{align}
\mathcal{H}(x)&=\pi\partial_{t}\varphi^{*}+\pi^{*}\partial_{t}\varphi-\mathcal{L}\nonumber\\
&=\frac{\hbar^{2}}{2m}\partial_{i}\varphi^{*}\partial^{i}\varphi+\frac{mc^{2}}{2}\st^{00}\varphi^{*}\varphi
\end{align}
leading, after integration by parts, to the Hamiltonian 
\begin{align}
&H\!=\!
\!\!\int\!\!\! d^{3}{x}\,\varphi^{*}\!(\ve{x},t)\!\Big(\!\!-\!\frac{\hbar^{2}}{2m}\partial_{i}\partial^{i}\!\!
+\!\frac{mc^{2}}{2}\st^{\!00}(\ve{x},t)\!\Big)\varphi(\ve{x},t)\label{eq:H final grav}
\end{align}
Promoting the field $\varphi(x)$ $(\varphi^{*}(x))$ and its conjugate momenta $\pi(x)$ $(\pi(x)^{*})$ to operators:
\begin{align}
\varphi(\ve{x},t)&\to \hat{\varphi}(\ve{x},t), \nonumber\\
\pi(\ve{x},t) &\to \hat{\pi}(\ve{x},t)
\end{align}
 and imposing the canonical quantization rule, \textit{i.e.}
\begin{align}
\com{\hat{\varphi}(\ve{x},t)}{\hat{\pi}(\ve{x},t)}&= \com{\hat{\varphi}^{\dagger}(\ve{x},t)}{\hat{\pi}^{\dagger}(\ve{x},t)}\nonumber\\
&=\ i\hbar\delta(\ve{x}-\ve{y})
\end{align}
one obtains the hamiltonian
\begin{align}
\hat{H}=  \int d^{3}x \hat{\varphi}^{\dagger}(\ve{x})H_{1}(x)\hat{\varphi}(\ve{x})
\end{align}
where 
\begin{align}
H_{1}(\ve{x},t)= -\frac{\hbar^{2}}{2m}\partial_{i}\partial^{i}+\frac{mc^{2}}{2}\st^{00}(\ve{x},t)
\end{align}
is the single particle hamiltonian expessed in the position basis.

\section{Stochastic Schr\"odinger Equation and non-faster than lightt signaling}\label{app:deriv}
The calculations leading to the main result of this paper are rather involved. 
In this appendix we provide the technical details of the derivation of Eq.~(12) and Eq.~(13).
The perturbation expansion obtained by combining Eq.~\eqref{eq:stocmast}  with Eq.~(9) and Eq.~(10) gives the
following system of equations 
\begin{widetext}
\begin{align}
i\hbar\partial_{t}\rho_{0,t}^{\st}= & \hat{H}_{0}\rho_{0,t}^{\st}-\text{H.c.}\nonumber \\
i\hbar\partial_{t}\rho_{1,t}^{\st}= & \hat{H}_{0}\rho_{1,t}^{\st}+\left(\sum_{i=1}^{N}(\hat{A}_{i}\st_{i}(t)-i\braket{A_{i}}_{t}^{0}\st_{i}^{\I}(t)+\hat{O}_{1}+\frac{1}{2}\braket{\hat{O}_{1}-\hat{O}_{1}^{\dagger}}_{t}^{0}\right)\rho_{0,t}^{\st}-\text{H.c.}\nonumber \\
i\hbar\partial_{t}\rho_{2,t}^{\st}= & \hat{H}_{0}\rho_{2,t}^{\st}+\left(\sum_{i=1}^{N}(\hat{A}_{i}\st_{i}(t)-i\braket{A_{i}}_{t}^{0}\st_{i}^{\I}(t)+\hat{O}_{1}-\frac{1}{2}\braket{\hat{O}_{1}-\hat{O}_{1}^{\dagger}}_{t}^{0}\right)\rho_{1,t}^{\st}\nonumber \\
 & -\left(\sum_{i=1}^{N}i\braket{A_{i}}_{t}^{1}\st_{i}^{\I}(t)-\hat{O}_{2}+\frac{1}{2}\braket{\hat{O}_{1}-\hat{O}_{1}^{\dagger}}_{t}^{1}+\frac{1}{2}\braket{\hat{O}_{2}-\hat{O}_{2}^{\dagger}}_{t}^{0}\right)\rho_{0,t}^{\st}-\text{H.c.}
\end{align}
where $\braket{A}_{t}^{n}=\Tr(\hat{A}\rho_{n,t}^{\st})$, and similarly for the other operators. We can formally
solve the above system of equations as follows:
\begin{align}\label{eq:pertordd}
\rho_{0,t}^{\st}= & e^{i\hat{H}_{0}t}\rho_{0}e^{-i\hat{H}_{0}t}\nonumber \\
\rho_{1,t}^{\st}= & -\frac{i}{\hbar}\sum_{i=1}^{N}\int_{0}^{t}d\tau\left(\hat{A}_{i}(\tau-t)\st_{i}(\tau)-i\braket{A_{i}}_{\tau}^{0}\st_{i}^{\I}(\tau)+\hat{O}_{1}(\tau-t)-\frac{1}{2}\braket{O_{1}-O_{1}^{\dagger}}_{\tau}^{0}\right)\rho_{0,t}^{\st}+\text{H.c.}\nonumber \\
\rho_{2,t}^{\st}= & -\frac{i}{\hbar}\sum_{i=1}^{N}\int_{0}^{t}d\tau\left(\hat{A}_{i}(\tau-t)\st_{i}(\tau)-i\braket{A_{i}}_{\tau}^{0}\st_{i}^{\I}(\tau)+\hat{O}_{1}(\tau-t)-\frac{1}{2}\braket{O_{1}-O_{1}^{\dagger}}_{\tau}^{0}\right)e^{i\hat{H}(t-\tau)}\rho_{1,\tau}^{\st}e^{-i\hat{H}(t-\tau)}\nonumber \\
 & -\frac{i}{\hbar}\sum_{i=1}^{N}\int_{0}^{t}d\tau\left(i\braket{A_{i}}_{\tau}^{1}\st_{i}^{\I}(\tau)-\hat{O}_{2}(\tau-t)+\frac{1}{2}\braket{\hat{O}_{1}-\hat{O}_{1}^{\dagger}}_{\tau}^{1}+\frac{1}{2}\braket{{O}_{2}-{O}_{2}^{\dagger}}_{\tau}^{0}\right)\rho_{0,t}^{\st}+\text{H.c.}\nonumber \\
\end{align}
where $\hat{A}_{i}(t)$ is the operator $\hat{A}_{i}$ in the interaction
picture at time $t$: 
\begin{align}
\hat{A}_{i}(t)=e^{\frac{i}{\hbar}\hat{H}_{0}t}\hat{A}_{i}e^{-\frac{i}{\hbar}\hat{H}_{0}t},
\end{align}
and similarly for the operator $\hat{O}$.
Now we are in the position to compute a closed equation for the averaged
density matrix $\av{\rho_{t}^{\st}}$. We plug the solutions in Eq.~\eqref{eq:pertordd}
into Eq.~\eqref{eq:pertorddzero}; in this way the stochasticity
is entirely contained in polynomials of $h$, whose correlations are
known. We can then explicitly compute the stochastic average of each
term. Collecting all pieces together, we arrive at the following perturbative equations for the ensemble,
which are valid up to order $\xi^{2}$:
\begin{align}\label{eq:masterapprox}
i\hbar\partial_{t}\rho_{0,t}= & \hat{H}_{0}\rho_{0,t}-\text{H.c.}\nonumber \\
i\hbar\partial_{t}\rho_{1,t}= & \hat{O}_{1}\rho_{0,t}+\frac{1}{2}\braket{O_{1}-O_{1}^{\dagger}}_{t}^{0}\,\rho_{0,t}-\text{H.c.}\nonumber \\
i\hbar\partial_{t}\rho_{2,t}= & -\frac{i}{\hbar}\sum_{i,j=1}^{N}\int_{0}^{t}d\tau\,S_{ij}(t,\tau)(\hat{A}_{i}-\braket{A_{i}}_{t}^{0})(\hat{A}_{j}(\tau-t)-\braket{A_{j}(\tau-t)}_{t}^{0})\rho_{0,t}\nonumber \\
 & +\frac{i}{\hbar}\sum_{i,j=1}^{N}\int_{0}^{t}d\tau\,D_{ij}(t,\tau)(\hat{A}_{j}(\tau-t)-\braket{A_{j}(\tau-t)}_{t}^{0})\rho_{0,t}(t,\tau)(\hat{A}_{i}-\braket{A_{i}}_{t}^{0})\nonumber \\
 & +\frac{i}{\hbar}\sum_{i,j=1}^{N}\int_{0}^{t}d\tau\,(S_{ij}(t,\tau)-D_{ij}(t,\tau))\braket{A_{i}(A_{j}(\tau-t)-\braket{A_{j}(\tau-t)}_{t}^{0})}_{t}^{0}\rho_{0,t}\nonumber \\
 & +\hat{O}_{2}+\frac{1}{2}\braket{O_{2}-{O}_{2}^{\dagger}}_{t}^{0}\rho_{0,t}-\text{H.c.}
\end{align}
The above equations are again non-linear. The non-linear terms can
be removed by choosing 
\begin{align}\label{Eq.Oseries}
\hat{O}_{1}= & 0\nonumber \\
\hat{O}_{2}= & +\frac{i}{\hbar}\sum_{i,j=1}^{N}\int_{0}^{t}d\tau(S_{ij}(t,\tau)-D_{ij}(t,\tau))\hat{A}_{i}(\hat{A}_{j}(t-\tau)-\braket{A_{j}(t-\tau)}_{t}^{0})\nonumber \\
 & -\frac{i}{\hbar}\sum_{i,j=1}^{N}\int_{0}^{t}d\tau(S_{ij}(t,\tau)-D_{ij}^{*}(t,\tau))\braket{A_{i}}_{t}^{0}\hat{A}_{j}(\tau-t)
\end{align}
Substituting this expression into Eq.~\eqref{eq:masterapprox} and
resumming the Taylor series one arrives at: 
\begin{align}
\partial_{t}\rho_{t}= & -\frac{i}{\hbar}{\hat{H}_{0}}{\rho_{t}}-\frac{\xi^{2}}{\hbar^{2}}\int_{0}^{t}d\tau\sum_{i,j=1}^{N}D_{ij}(t,\tau)\left(\hat{A}_{i}\hat{A}_{j}(\tau-t)\rho_{t}-\hat{A}_{j}(\tau-t)\rho_{t}\hat{A}_{i}\right)+\mathcal{O}(\xi^{3})+\text{H.c.}\nonumber \\
\end{align}
or equivalently:
\begin{align}
\begin{split}\partial_{t}{\rho}_{t}  &=-\frac{i}{\hbar}\com{\hat{H}_{0}}{\rho_{t}}-\frac{\xi^{2}}{\hbar^{2}}\sum_{i,j=1}^{N}\int_{0}^{t}d\tau\,D_{ij}^{\text{\tiny R}}(t,\tau)\com{\hat{A}_{i}}{\com{\hat{A}_{j}(\tau-t)}{\rho_{t}}}\nonumber\\
&-\frac{i\xi^{2}}{\hbar^{2}}\sum_{i,j=1}^{N}\int_{0}^{t}d\tau\,D_{ij}^{\I}(t,\tau)\com{\hat{A}_{i}}{\acom{\hat{A}_{j}(\tau-t)}{\rho_{t}}}+\mathcal{O}(\xi^{3}),
\end{split}
\label{mefinal1}
\end{align}
\end{widetext}

\section{Relation to the HPZ master equation}\label{HPZ}
Let us start with the center of mass master equation given by Eq.~(\ref{new8})
with the free particle Hamiltonian $\hat{H}_{0}^{\textrm{\tiny CM}}=\frac{\hat{\bm{P}}^{2}}{2m}$,
where $\hat{\bm{P}}$ is the center of mass momentum operator. 
We make two assumptions regarding the noise correlation functions $\tilde{D}^{\R}(\mathbf{Q},s)$,
$\tilde{D}^{\I}(\mathbf{Q},s)$ and on the center of mass state $\rho_{t}$.
Loosely speaking, we  restrict to a nearly Markovian regime and assume that
the exchanged momentum between noise and system is small. Mathematically,
we  give the sufficient conditions to expand the operators to quadratic
order, i.e. to order $\mathcal{O}(\hat{X}^{2})$, $\mathcal{O}(\hat{P}^{2})$,
$\mathcal{O}(\hat{X}\hat{P})$:
\begin{description}
\item [{(a)}] The noise correlation times are small and the state $\rho_{t}$
is such that: 
\begin{align}
e^{\frac{i}{\hbar}\hat{H}_{0}^{\textrm{\tiny CM}}s}\approx1+\frac{i}{\hbar}\hat{H}_{0}^{\textrm{\tiny CM}}s=1+\frac{i}{\hbar}\frac{\hat{P}^{2}}{2m}s.
\end{align}

\item [{(b)}] The noise momentum correlations are small and the state $\rho_{t}$
is such that: 
\begin{align}
e^{\frac{i}{\hbar}\mathbf{Q}\cdot\hat{\bm{X}}}\approx1+\frac{i}{\hbar}\mathbf{Q}\cdot\hat{\bm{X}}-\frac{1}{\hbar^{2}}(\mathbf{Q}\cdot\hat{\bm{X}})^{2}
\end{align}
Moreover, the noise momentum correlations depend only on the modulus $Q=|{\bf Q}|$: 
\begin{align}
&\tilde{D}^{\R}(\mathbf{Q},s)=\tilde{D}^{\R}(Q,s),\nonumber\\
&\tilde{D}^{\I}(\mathbf{Q},s)=\tilde{D}^{\I}(Q,s).
\end{align}
which is equivalent, as follows from Eq.~(\ref{new4}), to assuming a noise correlation isotropic in space ${D}^{\R}(\mathbf{r},s)={D}^{\R}(r,s)$ and ${D}^{\I}(\mathbf{r},s)={D}^{\I}(r,s)$ with $r=|\ve{r}|$.

\item [{(c)}] The noise correlation time $\tau_0$ is small with respect to the evolution time $t$.

\end{description}
Applying the above assumptions (a), (b) and (c), using the formula in Eq.~\eqref{new10} for the amplification factors and the identity
\begin{align}
&\int d\mathbf{Q}f(Q)\mathbf{(Q}\cdot\mathbf{X})(\mathbf{\mathbf{Q}}\cdot\mathbf{Y})=\nonumber\\
&=\left(\int_{0}^{\infty}dQf(Q)Q^{4}\right)\frac{4\pi}{3}\mathbf{X}\cdot\mathbf{Y},
\end{align}
where $f(Q)$ denotes a generic function, we can perform the $\mathbf{\mathbf{Q}}$
integration in Eq.~(\ref{new8}). After some algebra we
obtain the simplified master equation:
\begin{alignat}{1}\label{eq:m_2}
\frac{d{\rho}^{\text{\tiny CM}}_{t}}{dt}= & -\frac{i}{\hbar}\sum_{j=1}^{3}\left[\frac{\hat{P_{j}}^{2}}{2m},\rho_{t}\right]\nonumber\\
&-\eta\frac{A^{\R}(r_{C}^{\R})}{m_{0}^{2}}\sum_{j=1}^{3}\left[\hat{X}_{j},\left[\hat{X}_{j},{\rho}_{t}\right]\right]
\nonumber\\
 & +\varPi\frac{A^{\R}(r_{C}^{\R})}{m_{0}^{2}}\sum_{j=1}^{3}\left[\hat{X}_{j},\left[\hat{P_{j}}/m,\hat{\rho}\right]\right]\nonumber\\
 &-i\varUpsilon\frac{A^{\I}(r_{C}^{\I})}{m_{0}^{2}}\sum_{j=1}^{3}\left[\hat{X}_{j},\left\{ \hat{P_{j}}/m,{\rho}_{t}\right\} \right],
\end{alignat}
where 
\begin{align}
\eta & =\frac{m_{0}^{2}c^4\xi^{2}}{6\pi^{2}\hbar^{7}}\int_{0}^{\infty}dQ\int_{0}^{\infty}d\tau  \tilde{D}^{\R}(Q,\tau)Q^{4}, \label{eta}\\
\varPi & =\frac{m_{0}^{2}c^4\xi^{2}}{6\pi^{2}\hbar^{7}}\int_{0}^{\infty}dQ\int_{0}^{\infty}d\tau\,  \tau \tilde{D}^{\R}(Q,\tau)Q^{4},\\
\varUpsilon & =-\frac{m_{0}^{2}c^4\xi^{2}}{6\pi^{2}\hbar^{7}}\int_{0}^{\infty}dQ\int_{0}^{\infty}d\tau\, \tau \tilde{D}^{\I}(Q,\tau)Q^{4}
\end{align}
are three phenomenological parameters [given assumption (c), these do
not depend on $t$], while $A^{\R}(r_{C}^{\R})/m_{0}^{2}$ and
$A^{\I}(r_{C}^{\I})/m_{0}^{2}$ are dimensionless amplification factors, related
to the size and shape of the composite object as well as to the noise spatial correlation cutoffs $r_{C}^{\R}$ and $r_{C}^{\I}$  (see Sec. \ref{sec:Master-Equation}).\newline\newline
\indent Equation~(\ref{eq:m_2}) has the same structure as the HPZ master equation~\cite{hu1992quantum}, 
except for the absence of the HPZ term that breaks translational invariance.
The reason why the HPZ master equation breaks translational invariance
lies in its founding assumption: a particle in a harmonic potential coupled to
a bath of  oscillators. In our case, loosely speaking, the external
oscillators correspond to the complex noise, while the harmonic potential,
which explicitly breaks translational invariance, is absent. 
Our noise does not break translational invariance, as we have assumed explicitly that the correlation function is translationally invariant
(see Sec. \ref{sec:Master-Equation}).  
\bibliographystyle{unsrt}
\bibliography{Bibliografia} 
\end{document}